\documentclass[12pt]{article}

\usepackage{amssymb,amsmath,graphicx,setspace,color}

\usepackage{bibunits}

\makeatletter
\newcounter{mybibcounter}
\renewenvironment{thebibliography}[1]
     {\section*{\refname}%
      \@mkboth{\MakeUppercase\refname}{\MakeUppercase\refname}%
      \list{\@biblabel{\@arabic\c@mybibcounter}}%
           {\settowidth\labelwidth{\@biblabel{#1}}%
            \leftmargin\labelwidth
            \advance\leftmargin\labelsep
            \@openbib@code
            \@nmbrlisttrue\def\@listctr{mybibcounter}%
            \let\p@mybibcounter\@empty
            \renewcommand\themybibcounter{\@arabic\c@mybibcounter}}%
      \sloppy
      \clubpenalty4000
      \@clubpenalty \clubpenalty
      \widowpenalty4000%
      \sfcode`\.\@m}
     {\def\@noitemerr
       {\@latex@warning{Empty `thebibliography' environment}}%
      \endlist}
\makeatother

\usepackage[hyphens]{url}

\usepackage[nodayofweek]{datetime}
\newdateformat{mydate}{\twodigit{\THEDAY}{ }\monthname[\THEMONTH] \THEYEAR}

\usepackage{authblk}

\usepackage[nosort]{cite}

\usepackage[bf,skip=0pt,small]{caption}
\DeclareCaptionLabelSeparator{vert}{ $\boldsymbol{\vert}$ }
\usepackage{titlesec}
\titleformat{\section}{\fontsize{12}{15}\usefont{T1}{phv}{b}{n}\selectfont}{\thesection}{1em}{}
\titlespacing{\section}{0pt}{\parskip}{-\parskip}
\titlespacing{\subsection}{0pt}{\parskip}{-\parskip}
\titlespacing{\subsubsection}{0pt}{\parskip}{-\parskip}

\usepackage[left=1.5cm,top=1.5cm,right=1.5cm,bottom=1.5cm,nohead,nofoot]{geometry}

\newcommand{\sm}{$\sim$}
\newcommand{\dg}{$^{\circ}$}
\newcommand{\ha}{H$\alpha$}
\newcommand{\har}{H$\alpha + 0.6$}
\newcommand{\harr}{H$\alpha + 1.0$}

\newcommand\aapr{Astron. Astrophys. Rev.}

\newcommand\apjl{Astrophys. J. Lett.}
\newcommand\aap{Astron. Astrophys.}
\newcommand\apj{Astrophys. J.}
\newcommand\solphys{Sol.~Phys.}

\newcommand\nat{Nature}
\newcommand\pasj{Publ. Astron. Soc. Jpn.}

\def\mathbi#1{\textbf{\em #1}}

\makeatletter
\renewcommand{\maketitle}{\bgroup\setlength{\parindent}{0pt}
\begin{flushleft}	
  {\vskip-2mm \fontsize{12}{12}\usefont{T1}{phv}{m}{n}\selectfont \textcolor{blue}{Published in \textit{Nature Astronomy} 1, 0085 (2017) \url{http://www.nature.com/articles/s41550-017-0085}} \vskip 4mm}
  {\vskip-5.5mm \rule{\textwidth}{5pt} \vskip 2mm}
  \@title

  \bigskip
  \@author
\end{flushleft}\egroup
}

\renewcommand\@biblabel[1]{#1.}
\def\@cite#1#2{$^{\mbox{\scriptsize #1\if@tempswa , #2\fi}}$}

\let\oldbibliography\thebibliography
\renewcommand{\thebibliography}[1]{
	\oldbibliography{#1}
	\setlength{\itemsep}{-3pt}
}
\makeatother

\begin{document}

\title{{\fontsize{23}{26}\usefont{T1}{phv}{b}{n}\selectfont High-resolution observations of flare precursors in the low solar atmosphere}}

\author[1,2,3]{Haimin Wang}
\author[1,2,3]{Chang Liu}
\author[2]{Kwangsu Ahn}
\author[1,2,3]{Yan Xu}
\author[1,2,3]{Ju Jing}
\author[1,2,3]{Na Deng}
\author[1,2,3]{\\Nengyi Huang}
\author[4,5]{Rui Liu}
\author[6]{Kanya Kusano}
\author[3]{Gregory D. Fleishman}
\author[3]{Dale E. Gary}
\author[2,3]{\\Wenda Cao}

\affil[1]{ Space Weather Research Laboratory, New Jersey Institute of Technology, University Heights, Newark, NJ 07102-1982, USA}
\affil[2]{ Big Bear Solar Observatory, New Jersey Institute of Technology, 40386 North Shore Lane, Big Bear City, CA 92314-9672, USA}
\affil[3]{ Center for Solar-Terrestrial Research, New Jersey Institute of Technology, University Heights, Newark, NJ 07102-1982, USA}
\affil[4]{ CAS Key Laboratory of Geospace Environment, Department of Geophysics and Planetary Sciences, University of Science and Technology of China, Hefei 230026, China}
\affil[5]{ Collaborative Innovation Center of Astronautical Science and Technology, Hefei, China}
\affil[6]{ Institute for Space-Earth Environmental Research, Nagoya University, Furo-cho, Chikusa-ku, Nagoya, 464-8601, Japan}

\maketitle

\begin{bibunit}

\noindent {\fontsize{11}{11}\usefont{T1}{phv}{b}{n}\selectfont Solar flares are generally believed to be powered by free magnetic energy stored in the corona\cite{priest02}, but the build up of coronal energy alone may be insufficient for the imminent flare occurrence\cite{benz08}. The flare onset mechanism is a critical but less understood problem, insights into which could be gained from small-scale energy releases known as precursors, which are observed as small pre-flare brightenings in various wavelengths (e.g., refs\cite{bumba59,martin80,van86,kai83,warren01,asai06,chifor07,battaglia09,altyntsev12,fleishman15,zhangy15}), and also from certain small-scale magnetic configurations such as the opposite polarity fluxes\cite{kusano12,toriumi13,bamba13}, where magnetic orientation of small bipoles is opposite to that of the ambient main polarities. However, high-resolution observations of flare precursors together with the associated photospheric magnetic field dynamics are lacking. Here we study precursors of a flare using unprecedented spatiotemporal resolution of the 1.6~m New Solar Telescope, complemented by novel microwave data. Two episodes of precursor brightenings are initiated at a small-scale magnetic channel\cite{zirin93,kubo07,wang08,lim10} (a form of opposite polarity fluxes) with multiple polarity inversions and enhanced magnetic fluxes and currents, lying near the footpoints of sheared magnetic loops. The low-atmospheric origin of these precursor emissions is corroborated by microwave spectra. We propose that the emerging magnetic channel field interacts with the sheared arcades to cause precursor brightenings at the main flare core region. These high-resolution results provide evidence of low-atmospheric small-scale energy release and possible relationship to the onset of the main flare.}

We study the 22 June 2015 M6.5 flare (SOL2015-06-22T18:23) using \ha\ (line-center and red-wing) images and photospheric vector magnetograms obtained by the recently commissioned 1.6~m New Solar Telescope (NST)\cite{goode10,cao10} at Big Bear Solar Observatory (BBSO), which is stabilized by a high-order adaptive optics (AO) system (see Methods). In particular, the vector field data is taken by the Near InfraRed Imaging Spectropolarimeter (NIRIS)\cite{cao12} at the 1.56$\mu$ Fe~{\sc I} line. These observations have the highest spatial resolution ever achieved for the solar observations (\sm70~km for \ha\ and \sm170~km for vector field) and rapid cadence (28~s for \ha\ and 87~s for vector field). Also used are flare microwave spectra and time profiles from the new Expanded Owens Valley Solar Array (EOVSA; see Methods), and time profiles of hard X-ray (HXR) and soft X-ray (SXR) fluxes from the Reuven Ramaty High Energy Solar Spectroscopic Imager (RHESSI)\cite{lin02} and the Geostationary Operational Environmental Satellite (GOES)-15, respectively. Ancillary  data of full-disk corona images and magnetograms from the Solar Dynamics Observatory (SDO)\cite{pesnell12} are additionally used.

\begin{figure*}[!t]
\centering
\includegraphics[width=7.38in]{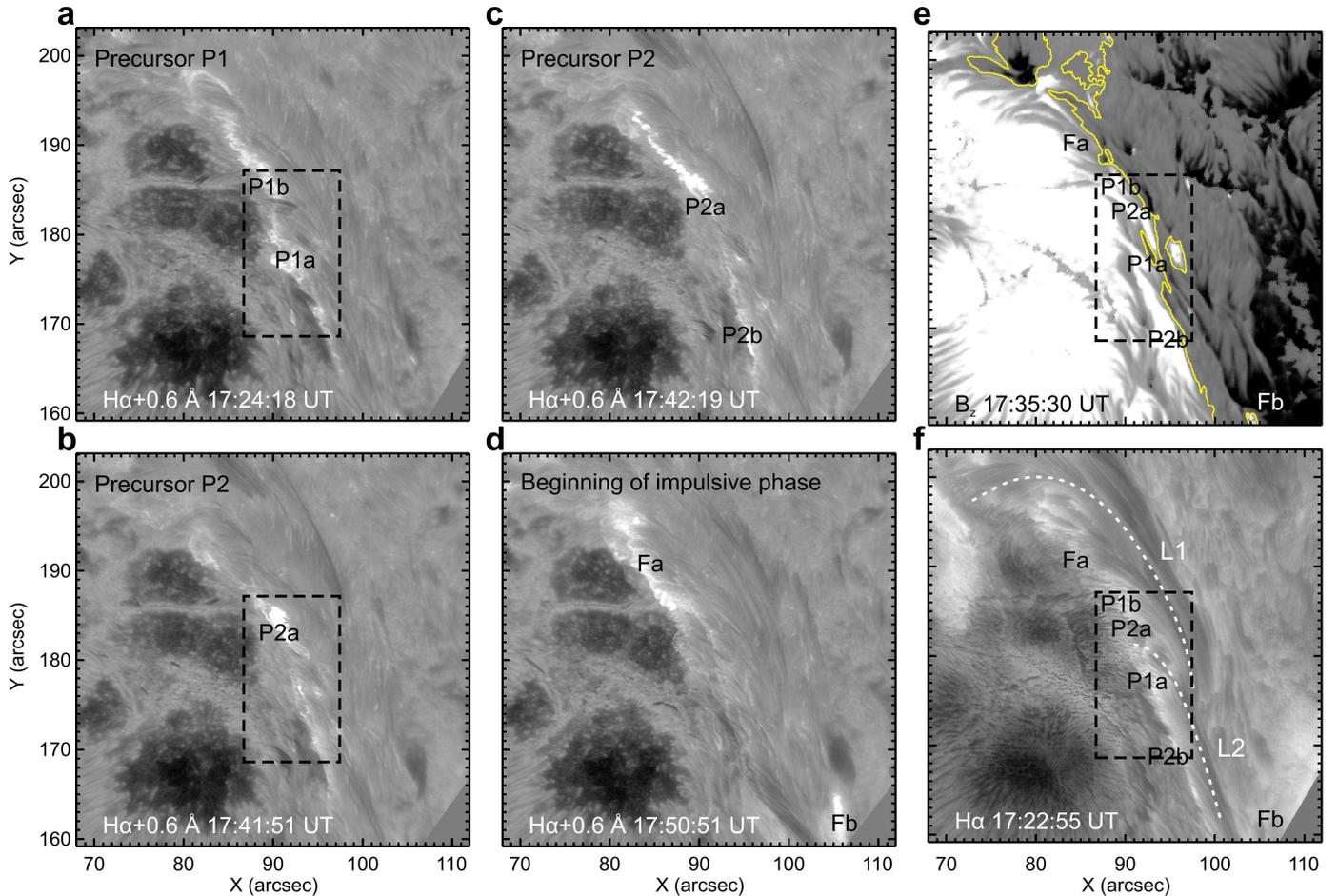}
\vskip -2mm
\caption{\textbf{Precursor brightenings.} BBSO/NST chromospheric \har~\AA\ (\textbf{a}-\textbf{d}) and \ha\ center-line (\textbf{f}) images (in logarithmic scale) in comparison with NIRIS photospheric vertical magnetic field $B_z$ (\textbf{e}; scaled between $\pm$1500 G), showing the core region structure of the 22 June 2015 M6.5 flare. Brightenings labeled P1a/P1b, P2a/P2b, and Fa/Fb appear during the precursor episodes P1, P2, and at the beginning of the flare impulsive phase, respectively, the timing of which are indicated in Supplementary Fig.~1. The dashed box in \textbf{a}, \textbf{b}, \textbf{e}, and \textbf{f} denote the field of view (FOV) of Figs~2b and 3a-c. The yellow contour in \textbf{e} indicates the PIL. The dashed lines in \textbf{f} illustrate sheared arcade loops L1 and L2 (also see Supplementary Fig.~2e). All the images were registered with respect to 22 June 2015 17:24 UT. \label{f1}}
\end{figure*}

The long-duration 22 June 2015 M6.5 flare occurred near the disk center (8$^{\circ}$W, 12$^{\circ}$N) at NOAA active region (AR) 12371. Time profiles of flare emissions in different wavelengths (including HXR, SXR, and microwave) clearly display that soon before the flare impulsive phase starting from \sm17:51~UT, there are two short episodes of smaller magnitude emissions at \sm17:24~UT and \sm17:42~UT, which we denote as P1 and P2, respectively (see Supplementary Fig.~1). It is found that these emissions are only possible to stem from the AR of the imminent M6.5 flare, and that simultaneous \ha\ brightenings are observed with NST in the flaring core region (see Supplementary Video 1). Thus they can be regarded as precursors of the M6.5 flare. Fine structural evolution of the associated precursor brightenings in NST \ha\ and the surface magnetic structure are presented in Fig.~1. Specifically, the brightening associated with the precursor episode P1 first appears as a kernel P1a in NST at 17:23:21~UT (also discernible in UV/EUV; see Supplementary Fig.~2a,d), then quickly turns into an elongated structure with another kernel P1b (Fig.~1a and Supplementary Fig.~2b). Later from \sm17:38~UT, fine-scale brightening starts to be seen in the south and travels northeastward apparently along the previously brightened regions in P1. The precursor episode P2 starts around 17:42~UT as a kernel P2a (Fig.~1b), then another P2b is formed in the south (Fig.~1c). All the above brightenings exhibit little propagation towards the east. From \sm17:46~UT one of the main flare ribbon Fa seems to develop from the P2a region (see Fig.~1d and Supplementary Video 1).

Regarding the precursor brightenings, we notice that P1a/P1b and P2a/P2b all lie along a narrow lane of largely positive magnetic polarity, a few arcseconds to the east of the magnetic polarity inversion line (PIL) (see Fig.~1e). The region of P1a is co-spatial with small-scale mixed polarities (discussed below) located near the footpoints of large-scale sheared arcades, which are approximately illustrated as L1 and L2 in \ha\ and 193~\AA\ (see Fig.~1f and Supplementary Fig.~2d,e). Some other brightenings are seen near the southern footpoint of L2 in the negative field region (e.g., see Supplementary Fig.~2a-d). Furthermore, precursor brightenings predominantly move along (parallel to) the PIL, in contrast to the separation motion of the main flare ribbons Fa and Fb (see Fig.~1d, Supplementary Fig.~2c, and Supplementary Video 1) away from the PIL that complies with the standard flare model\cite{kopp76}.

\begin{figure*}[!t]
\centering
\includegraphics[width=7.5in]{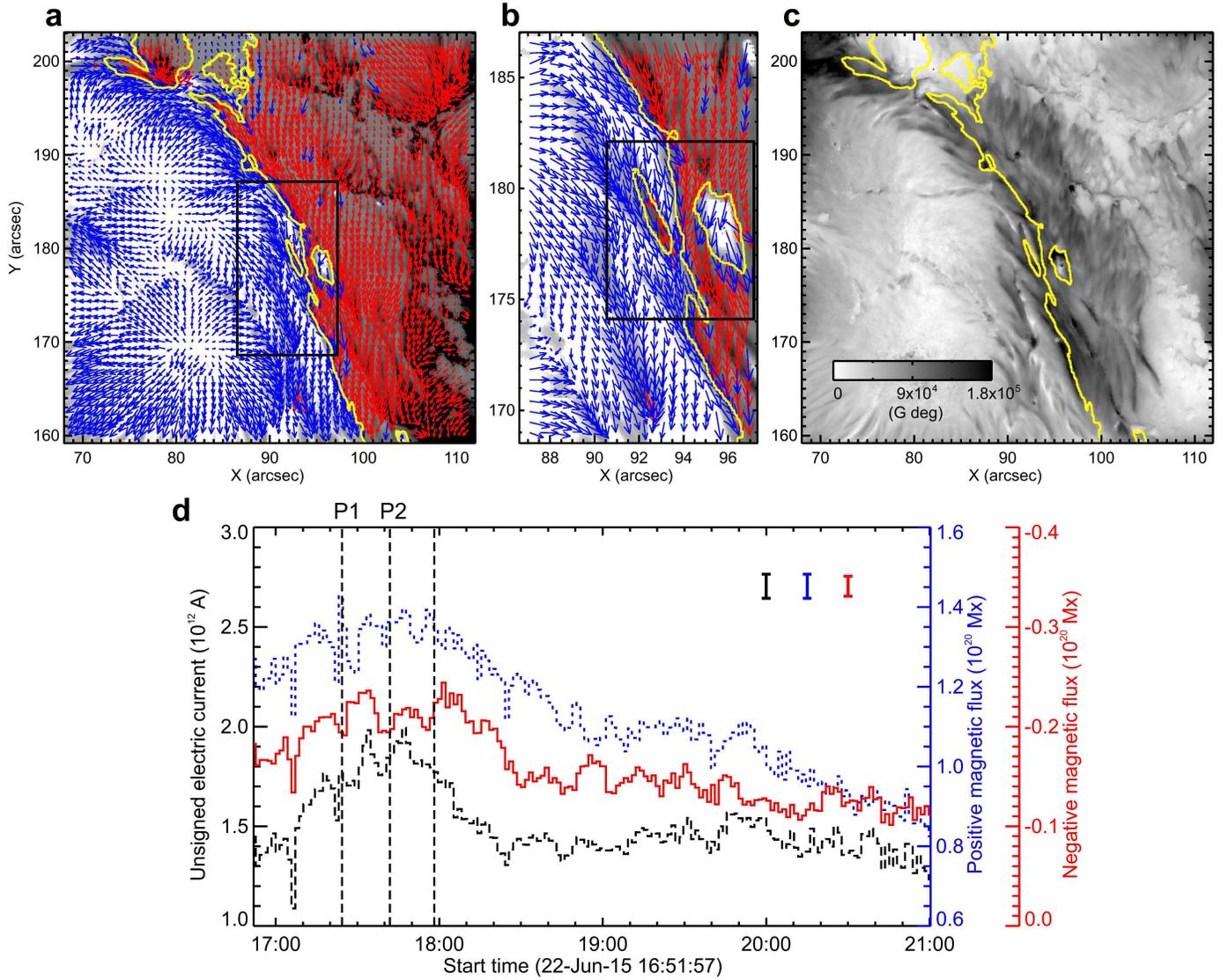}
\vskip 0mm
\caption{\textbf{Magnetic field structure and evolution.} (\textbf{a},\textbf{b}) BBSO/NST NIRIS photospheric vertical magnetic field (scaled between $\pm$1500~G) at 17:35:30~UT superimposed with arrows (for clarity, those in positive/negative fields are coded in blue/red) representing horizontal magnetic field vectors. The box in \textbf{a} (same as those in Fig.~1a,b,e,f) denotes the FOV of \textbf{b}, in which the magnetic channel structure can be obviously observed. (\textbf{c}) Distribution of magnetic shear (see Methods). The overplotted yellow contour in \textbf{a}--\textbf{c} is the PIL. (\textbf{d}) Temporal evolution of total positive (blue dotted line) and negative (red solid line) magnetic fluxes and the unsigned electric current (black dashed line; see Methods), calculated over the magnetic channel region enclosed by the box in \textbf{b}. The plotted representative error bars correspond to 1 s.d. calculated over the post-flare period 18:50 to 19:40~UT, demonstrating the background variation. The first two vertical dashed lines indicate the times of precursor episodes P1 and P2 at \sm17:24 and \sm17:42~UT, respectively; the third vertical dashed line denotes the peak time of the flare nonthermal emission in microwave at \sm17:58~UT (see Supplementary Fig.~1).\label{f2}}
\end{figure*}

\begin{figure*}[!t]
\centering
\includegraphics[width=7.35in]{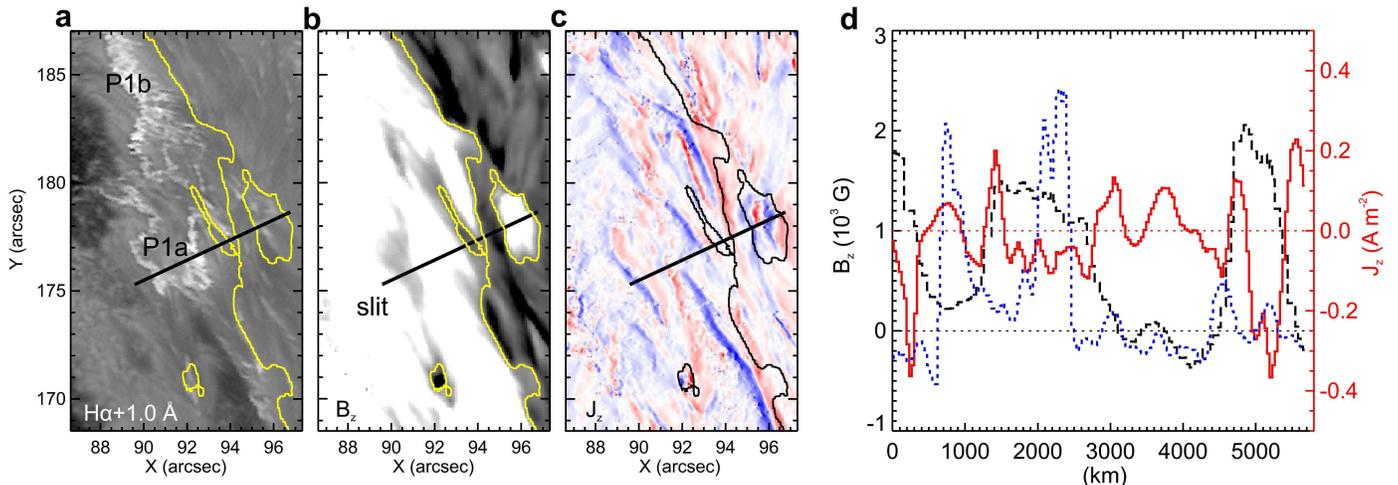}
\vskip 0mm
\caption{\textbf{Properties of magnetic channel region.} (\textbf{a}) \harr~\AA\ image at the time of the precursor episode P1. (\textbf{b},\textbf{c}) NIRIS vertical magnetic field $B_z$ and vertical current density $J_z$ derived from horizontal field (see Methods) at a preflare time. The shown FOV of \textbf{a}-\textbf{c} is denoted by the box in Figs~1a,b,e,f and 2a and is the same as that of Fig.~2b. The yellow (black) contour in \textbf{a} and \textbf{b} (\textbf{c}) is the PIL. (\textbf{d}) Profiles of $J_z$ (red solid line), $B_z$ (black dashed line), and normalized \harr~\AA\ intensity (blue dotted line; in arbitrary unit) along a slit (indicated by the thick black line in \textbf{a}-\textbf{c}) through the magnetic channel. The distance in the abscissa is set to zero at the southeastern end of the slit. Error bars are omitted as the uncertainties are too small compared to the axis scale. \label{f3}}
\end{figure*}

In Fig.~2, we present the NIRIS vector magnetic field measurement of the flare core region. The magnetic field is highly sheared with respect to the PIL, especially in the precursor brightening region (Fig.~2a,b). This signifies a high degree of nonpotentiality, as reflected by the concentration of magnetic shear along the PIL (Fig.~2c; see equation (1) in Methods). In the region around the initial precursor brightening P1a (enclosed by the box in Fig.~2b), we observe elongated, alternating positive and negative polarities in a fine scale of \sm3,000~km, constituting a miniature version of the magnetic channel structure\cite{zirin93} also recognized as the opposite polarity (OP)-type field\cite{kusano12} (see also Supplementary Video 2). Importantly, both the negative and positive fluxes within the channel exhibit an increase (by \sm$-$6.6~$\times$~10$^{18}$ and \sm9.3~$\times$~10$^{18}$~Mx, respectively, in about half an hour) temporally associated with the occurrence of the precursor episodes P1 and P2, and a decrease after the peak of the flare nonthermal emission (Fig.~2d). These imply that the dynamic evolution of the magnetic channel region is closely related to the triggering and subsequent evolution of the flare. Finding the spatial and temporal correlation between the appearance of the precursor brightenings and the properties of the magnetic field structure and evolution is possible because of the high spatial and temporal resolution of the NIRIS data.

\begin{figure*}[!t]
\centering
\includegraphics[width=6.5in]{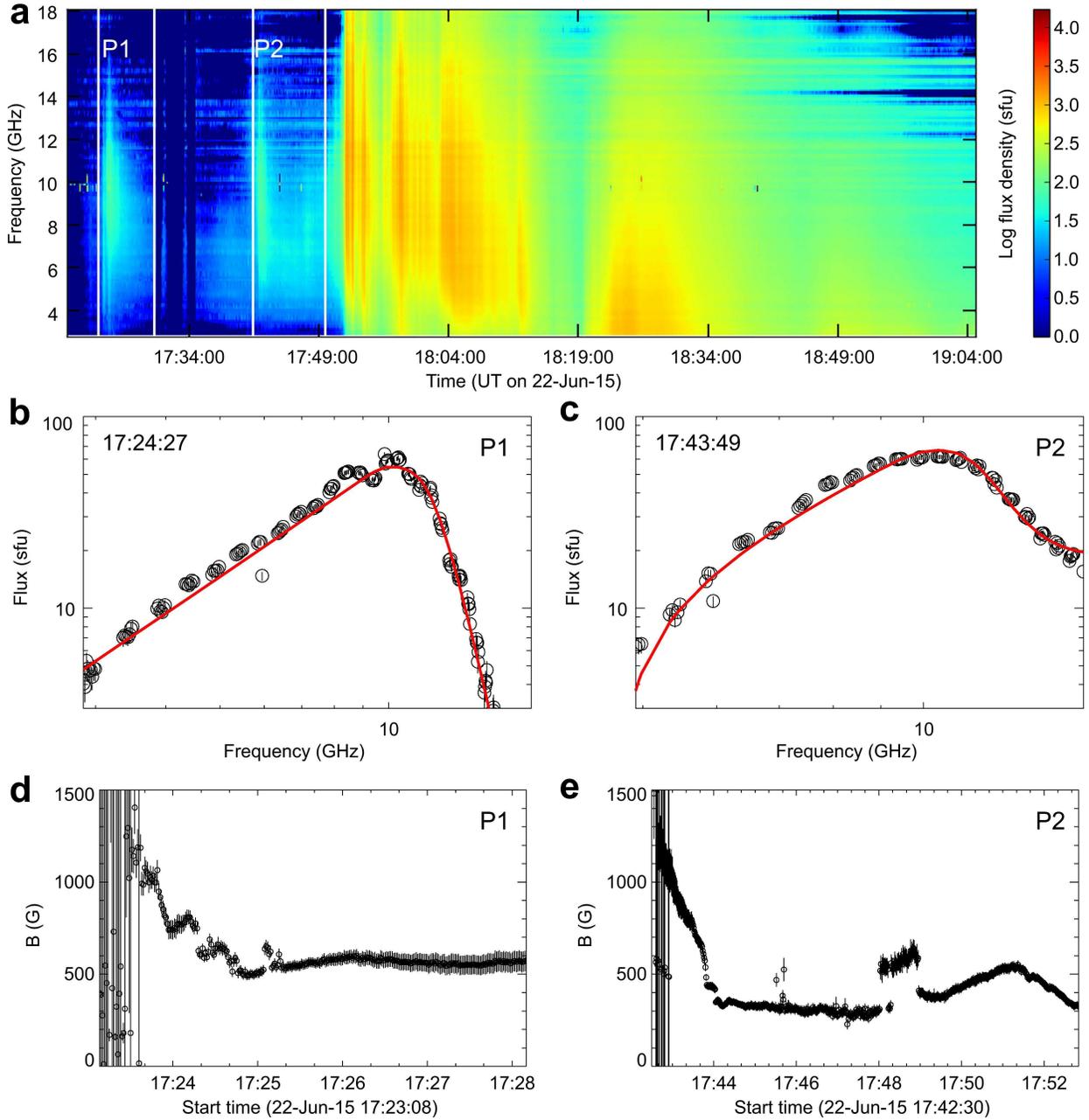}
\vskip 0mm
\caption{\textbf{Microwave emission.} (\textbf{a}) EOVSA total power dynamic spectra (with the preflare quiet Sun and AR contribution subtracted) of microwave emission, covering two precursor episodes P1 and P2 and the impulsive phase of the flare. The four white vertical stripes are data gaps. (\textbf{b},\textbf{c}) Sample spectral fittings at P1 and P2 using a gyrosynchrotron source function (see Methods). (\textbf{d},\textbf{e}) Evolution of the derived magnetic field strength in the two precursor periods. The error bars represent 1 s.d. statistical errors returned by the spectral fit. \label{f4}}
\end{figure*}

We further characterize the fine-scale properties of the magnetic channel region in Fig.~3. A comparison reveals that the precursor brightenings P1a/P1b  (Fig.~3a) appear in the vicinity of regions of strong vertical current density (curl of the horizontal field; calculated by equation (2) in Methods and shown in Fig.~3c), similar to what has been found previously for main flare ribbons (e.g., ref.\cite{janvier14}). We also place a slit right across the magnetic channel and plot the profiles of the vertical magnetic field and current density along it (black dashed and red solid lines in Fig.~3d). The results show 5 (11) time reversals of magnetic polarity (electric current) within \sm3000 (6000)~km, demonstrating the complexity of this channel region at small scales. The profile of \harr~\AA\ along the same slit (blue dotted line in Fig.~3d) shows that the brightening has a fine width of 500~km or less. Similar to the magnetic flux, the unsigned vertical electric current integrated over the channel region shows a clear increase (by \sm5~$\times$~10$^{11}$~A in about half an hour) closely related with the timing of precursors P1/P2 and also a decrease after the main energy release (Fig.~2d). Such a flare-related evolutionary pattern is not found in other areas within the observed FOV. These results are suggestive that the enhancing magnetic channel structure corresponds to an emerging small current-carrying flux tube\cite{lim10}, which might be dissipated via reconnection with ambient fields after the flare.

To extend the measurement of the magnetic field to the three-dimensional (3D) domain, we turn to the analysis of microwave observations. For reasonably uniform emission sources (such as flare precursors), the magnetic field can be derived from the total power data\cite{fleishman16}. In Fig.~4a the EOVSA total power dynamic spectrum (intensity recorded in a time-frequency diagram, averaged from the measurements of 20 dishes), with the preflare quiet Sun and AR contribution subtracted, shows microwave flaring emissions of the two precursor episodes P1/P2 and the impulsive phase. Our analyses show that the microwave spectra in the precursor periods can be modeled as  quasi-thermal, gyrosynchrotron emission sources (e.g., Fig.~4b,c; see Methods). Their instantaneous spectral shape is determined by several physical parameters (e.g., temperature,  magnetic field, and electron density)\cite{stahli89}. For our purpose, we only consider the temporal evolution of the magnetic field during the precursor periods, which is derived from the spectral fittings and plotted in Fig.~4d,e. The magnetic field strength of precursors P1 and P2 is strong ($\gtrsim$1000~G) at the beginning, then gradually decreases to a lower level (500 and 300~G, respectively). Based on the relationship between the average magnetic field strength and height above the surface as suggested by a nonlinear force-free field (NLFFF) extrapolation model of this flaring region (see Methods and Supplementary Fig.~3b), this result indicates that both precursor emissions are initially located in the low atmosphere (at photospheric/chromospheric level), which corroborates the NST analysis.

Taken together, BBSO/NST \ha\ images complemented by EOVSA microwave  observation identify the low-atmospheric precursor emissions in close relation to the onset of the main flare. We propose that the present event proceeds in a way consistent with the model of ref.\cite{kusano12}: an emerging small-scale flux tube, as signified by the strengthening magnetic channel\cite{lim10} of OP type\cite{kusano12}, interacts with and facilitates the reconnection of the ambient legs of large-scale sheared loops rooted in major polarities (Supplementary Fig.~2d), producing precursor brightenings. These sheared loops are also demonstrated by the NLFFF model (Supplementary Fig.~3a), in which they lie close to the surface with an apex height of \sm5~Mm. The reconnection site between the small-scale emerging flux and sheared arcades could be located at the photospheric/chromospheric level; subsequently, the accelerated particles can quickly propagate and cause brightenings in other remote footpoints (e.g., P1c in Supplementary Fig.~2a,b,d). During the two precursor periods, the motion of \ha\ brightening kernels along the PIL may reflect the successive reconnection of different branches of the sheared arcade loops, which eventually leads to the imminent eruption of the main flare (Supplementary Fig.~2f). We made no attempt to compare our observations with other flare triggering scenarios, as the model of ref.\cite{kusano12} is perhaps the only one that incorporates both the small- and large-scale magnetic structures (i.e., small OP-type field and overlying sheared arcades). As sheared arcade systems are often present in flaring ARs, more observations of the low solar atmosphere, especially high-resolution photospheric magnetic field measurements with a good temporal coverage, are desired to further tackle the problem of the flare onset mechanism, which is critically related to the space weather forecast.

\bigskip
\renewcommand\refname{References (Main)}


\end{bibunit}

\section*{Acknowledgements}
\noindent We thank the teams of BBSO, EOVSA, SDO, RHESSI, GOES, and Hinode in obtaining the data. This work was supported by NASA under grants NNX13AF76G, NNX13AG13G, NNX14AC12G, NNX14AC87G, NNX16AL67G, and NNX16AF72G, and by NSF under grants AGS 1250374, 1262772, 1250818, 1348513, 1408703, and 1539791. R.L. acknowledges the support from the Thousand Young Talents Program of China and NSFC 41474151. K.K. acknowledges the support from MEXT/JSPS KAKENHI 15H05814. The BBSO operation is supported by NJIT, US NSF AGS 1250818, and NASA NNX13AG14G grants, and  partly supported by the Korea Astronomy and Space Science Institute and Seoul National University and by the strategic priority research program of CAS with Grant No. XDB09000000.

\bigskip
\section*{Author contributions}
\noindent H.W. initiated the idea and carried out the data processing, analysis, interpretation, and manuscript writing. C.L. contributed to the azimuth disambiguation of NIRIS data, data analysis and interpretation, and manuscript revision. K.A. developed tools for NIRIS data calibration,  polarization inversion, and processed the NIRIS data.  Y. X. was the PI of this BBSO/NST observation run and contributed to the data processing. J.J. and N.D. contributed to the data analysis. N.H. contributed to this NST observation run. R.L. contributed to the NLFFF modeling and result interpretation. K.K contributed to the interpretation of observations. G.D.F. and D.E.G. carried out the microwave data analysis and modeling. W.C. developed instruments at BBSO. All authors discussed the results and commented on the manuscript.

\bigskip
\section*{Author information}
\noindent Correspondence and requests for materials should be addressed to H.W. (haimin.wang@njit.edu) or W.C. (wenda.cao@njit.edu).

\bigskip
\section*{Methods}

\begin{bibunit}

\noindent\textbf{Optical wavelength observations and reduction.} NST is a 1.6~m off-axis telescope at BBSO operated by New Jersey Institute of Technology. Currently, it produces highest spatial resolution (e.g., 70~km when observing around 6000~\AA), diffraction-limited observations of the Sun, aided by a 308-element AO system and speckle-masking image reconstruction. During the period of \sm16:50--23:00~UT on 22 June 2015, NST observes the M6.6 flare at NOAA AR 12371. The data taken include spectroscopic observations in H$\alpha$ line center and $\pm$0.6 and $\pm$1.0~\AA\ (with a bandpass of 0.07~\AA), and also images in the TiO band (a proxy for the continuum photosphere around 7,057~\AA). The images have a spatial resolution of \sm70--80~km and a cadence ranging from 15--28~s.

Notably, this study makes the first scientific use of the data from NIRIS, which measures the photospheric magnetic field during this observation run. This also makes NIRIS's very first coverage of a flare observed with NST. NIRIS utilizes dual Fabry-P{\'e}rot etalons that provide a 60,000~km~$\times$~60,000~km FOV and a great throughput over 90\%. NIRIS achieves a spatial resolution of \sm170~km at the 1.56$\mu$ Fe~{\sc i} line (with a bandpass of 0.1~\AA). The cadence of NIRIS data (full set of vector magnetograms) for this 22 June 2015 observation run is 87~s with intended delay. The 1.56$\mu$ line offers a high Lande $g$-factor of 3, which can help increase the signal strength of magnetograms. Although this line has a lower diffraction limit than some visible lines, it produces much more stable images under atmospheric turbulence.

For the NIRIS observations, first, the dual beam optical design images two simultaneous polarization states onto a 2,024~$\times$~2,048 HgCdTe closed-cycle IR array that undergoes a thermo-electric cooling. A combination of a linear polarizer and a quarter-wave plate is employed in the telescope structure to create pure polarization states, after which the responses of the following optical elements through the detector are measured. This approach is able to eliminate the crosstalk among the Stokes Q, U, and V.

Second, the NIRIS data undergoes Stokes inversion using the Milne-Eddington technique, through which several key physical parameters (including total magnetic field, azimuth angle, inclination, Doppler shift) can be extracted. For successful fittings with Milne-Eddington-simulated profiles, initial parameters are pre-calculated to be in proximity to the observed Stokes profiles\cite{ahn16}. The accuracy of the resulted vector field data is 10~G for the line-of-sight (LOS) component and 100~G for the transverse component.

For this 22 June 2015 observation, we validate the NIRIS data by checking against those obtained from two spaceborne instruments, the spectropolarimeter (SP) of Hinode's 0.5~m Solar Optical Telescope\cite{tsuneta08} and SDO's Helioseismic and Magnetic Imager (HMI)\cite{schou12}. Hinode/SP and SDO/HMI have a spatial resolution of about 230 and 725~km, and a temporal cadence of a few hours and 12 minutes, respectively. In Supplementary Fig.~4, we compare LOS field, transverse field, and the azimuthal angle from the three instruments measured around 22 June 2015 22:35~UT. It can be clearly seen that the magnetic structures observed by NIRIS, Hinode/SP, and SDO/HMI are highly similar, but more details are present in NIRIS data due to its higher spatial resolution. A cross-correlation analysis reveals that data acquired by NIRIS matches well with those taken by Hinode/SP and SDO/HMI, with a correlation coefficient of 0.97, 0.9, and 0.8 for the LOS field, transverse field, and the azimuthal angle measurements, respectively. This demonstrates the superiority of NIRIS observations in making high spatiotemporal resolution studies of photospheric magnetic field evolution.

Third, to properly explore the magnetic field structure, we further process the NIRIS vector magnetograms resulted from inversion to (1) resolve the 180\dg\ azimuthal ambiguity using the AZAM code\cite{leka09b}, which is based on the ``minimum energy'' ambiguity resolution method\cite{metcalf94,metcalf06}, and (2) remove the projection effects by transforming the observed vector fields (LOS and transverse) to those in heliographic coordinates (vertical $B_z$ and horizontal $B_x$ and $B_y$ fields) using the equations of ref.\cite{gary_hagyard90}. To characterize the nonpotentiality of the active region, we calculate magnetic shear $\tilde{S}$, defined as the product of field strength and shear angle\cite{wang94,wang06shear}:
\begin{equation}
\tilde{S}=|\mathbi{B}| \cdot \theta \ ,
\end{equation}
\noindent where $\theta={\rm cos}^{-1}(\mathbi{B} \cdot \mathbi{B}_p)/(BB_p)$, and the subscript $p$ represents the potential field (here computed using the Green's function method\cite{metcalf08}). We also derive the vertical current density $J_z$:
\begin{equation}
J_z = \frac{1}{\mu_0} \left( \nabla \times \mathbi{B}\right)_z = \frac{1}{\mu_0} \left( \frac{\partial B_y}{\partial x} - \frac{\partial B_x}{\partial y} \right) \ ,
\end{equation}
\noindent where $\mu_0$ is the permittivity of the vacuum.

\bigskip
\noindent\textbf{Context full-disk observations.} We complement high-resolution NST data with full-disk SDO observations for a better understanding of the overall picture of the flare. Images shown in Supplementary Fig.~2 include those from SDO's Atmospheric Imaging Assembly (AIA)\cite{lemen12} 1700~\AA\ (temperature minimum region and photosphere), 193~\AA\ (corona and hot flare plasma), and 94~\AA\ (hot flare plasma) passbands, and SDO's HMI.

\bigskip
\noindent\textbf{Microwave observation and analysis.} It is notable that this study also makes the first scientific use of the EOVSA data. EOVSA is a newly upgraded, solar-dedicated radio array consisting of 13 antennas of 2.1~m diameter, which are equipped with receivers designed to cover the 1--18~GHz frequency range. Two large (27~m diameter) dishes are being outfitted with He-cooled receivers for use in calibration of the small dishes\cite{gary16}. EOVSA has just started the scientific operation.

For the event under study, the EOVSA microwave observation covers both the precursors and the main flare. Although the microwave spectrum is broadband during the main flare (which is a result of source nonuniformity in space), the preflare phase demonstrates reasonably narrow spectra consistent with a quasi-uniform source. This kind of source can be conclusively forward-fit using the gyrosynchrotron source function to recover the physical parameters responsible for the observed spectral shape\cite{fleishman16}. Inspection of the detailed spectral shape at the preflare phase reveals that the spectra are almost thermal, which greatly simplifies the fitting and enhances the reliability of the fit results\cite{fleishman15}. In order to quantitatively define the gyrosynchrotron source function, we use the fast gyrosynchrotron codes developed by the ref.\cite{fleishman10} and apply the optimization scheme as described by the ref.\cite{fleishman09}. The fit returns evolution of the plasma density, temperature, and the magnetic field strength (as shown in Fig.4d,e). In this analysis, we focus on the evolution of magnetic field at the instantaneous location of the radio burst, which recovers a significant elevation of the radio source with time and, thus, allows us to extend the analysis to the 3D domain. The use of magnetic field extrapolation model (see below) maps the coronal magnetic field probed from the microwave data to the actual heights above the active region.

\bigskip
\noindent\textbf{Coronal magnetic field extrapolation.} In order to disclose the magnetic structure above the flaring active region, we use the NLFFF extrapolation technique to construct 3D magnetic field. As NIRIS data has a limited FOV, we carry out the magnetic field modeling based on the lower-resolution SDO/HMI vector data. A preprocessing procedure\cite{wiegelmann06} is first performed to minimize the net force and torque in the observed photospheric field. A ``weighted optimization'' method\cite{wheatland00,wiegelmann04} is then applied to derive the NLFFF, from which magnetic field lines are computed. Physical properties of field lines, such as the magnetic twist, can be further deduced\cite{liur16}.

\bigskip
\noindent\textbf{Data availability.} The data that support the plots within this paper and other findings of this study are available from the corresponding authors upon reasonable request.

\bigskip
\renewcommand\refname{References}


\end{bibunit}

\newpage

\begin{figure*}[!t]
\setcounter{figure}{0}
\makeatletter
\renewcommand{\fnum@figure}{Supplementary Figure~\thefigure}
\makeatother
\centering
\includegraphics[width=7in]{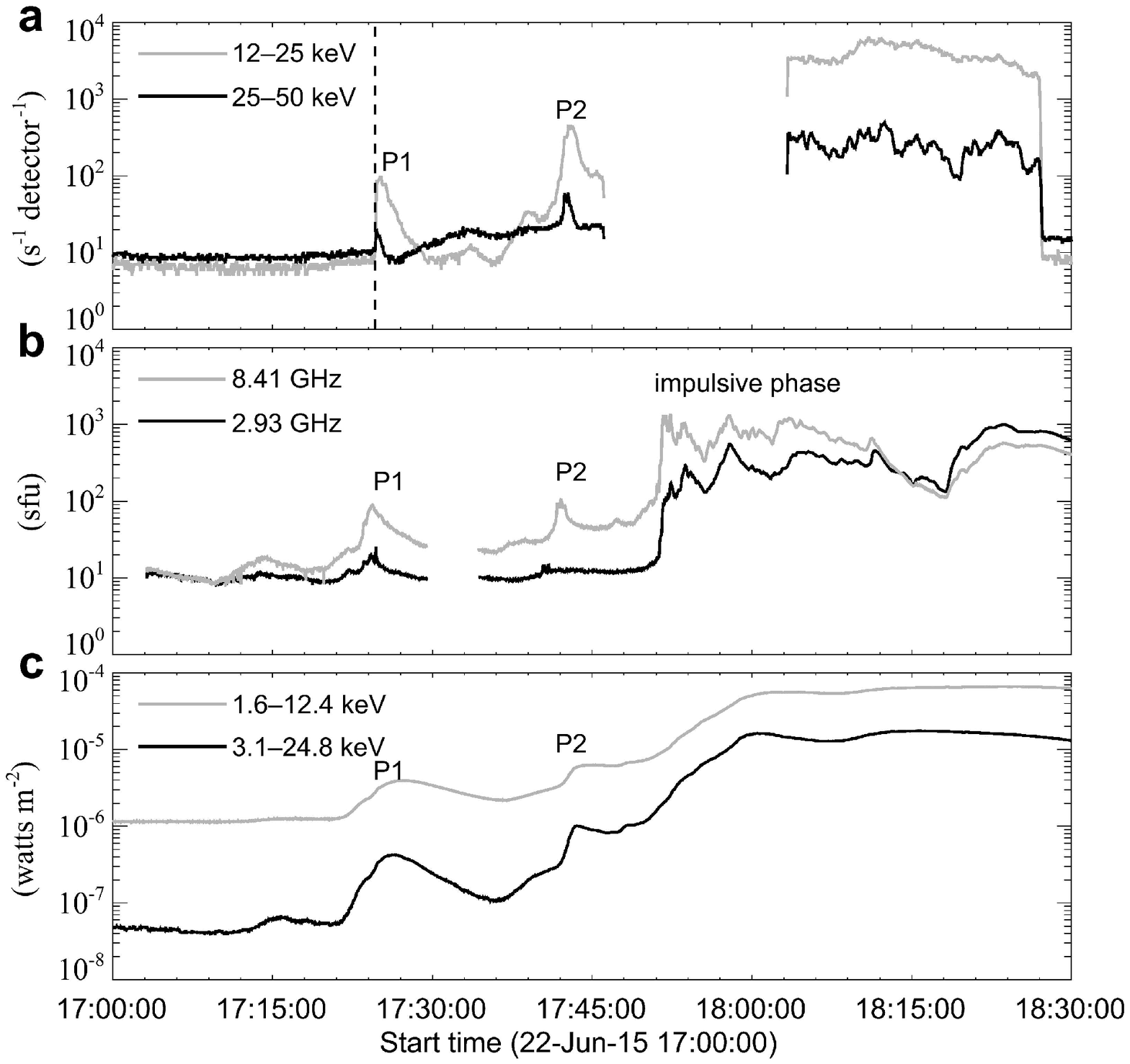}
\caption{\textbf{Flare light curves.} (\textbf{a}) RHESSI HXR corrected count rate in 12--25 and 25--50~keV energy bands. RHESSI was in night time between \sm16:53 and \sm17:24~UT (vertical dashed line), and had no observation during \sm17:46 and \sm18:03~UT due to South Atlantic Anomaly passage. (\textbf{b}) EOVSA microwave flux density at 2.93 and 8.41 GHz. (\textbf{c}) GOES SXR flux in 1.6--12.4 and 3.1--24.8~keV energy bands. P1 and P2 mark the two flare precursor episodes. A weak emission peak can also be seen in SXR and microwave at 17:15~UT; however, no brightening kernels in NST \ha\ and AIA 1700~\AA\ images are observed around this time. Instrument measurement errors are omitted. \label{sf1}}
\end{figure*}

\begin{figure*}[!t]
\makeatletter
\renewcommand{\fnum@figure}{Supplementary Figure~\thefigure}
\makeatother
\centering
\includegraphics[width=7.3in]{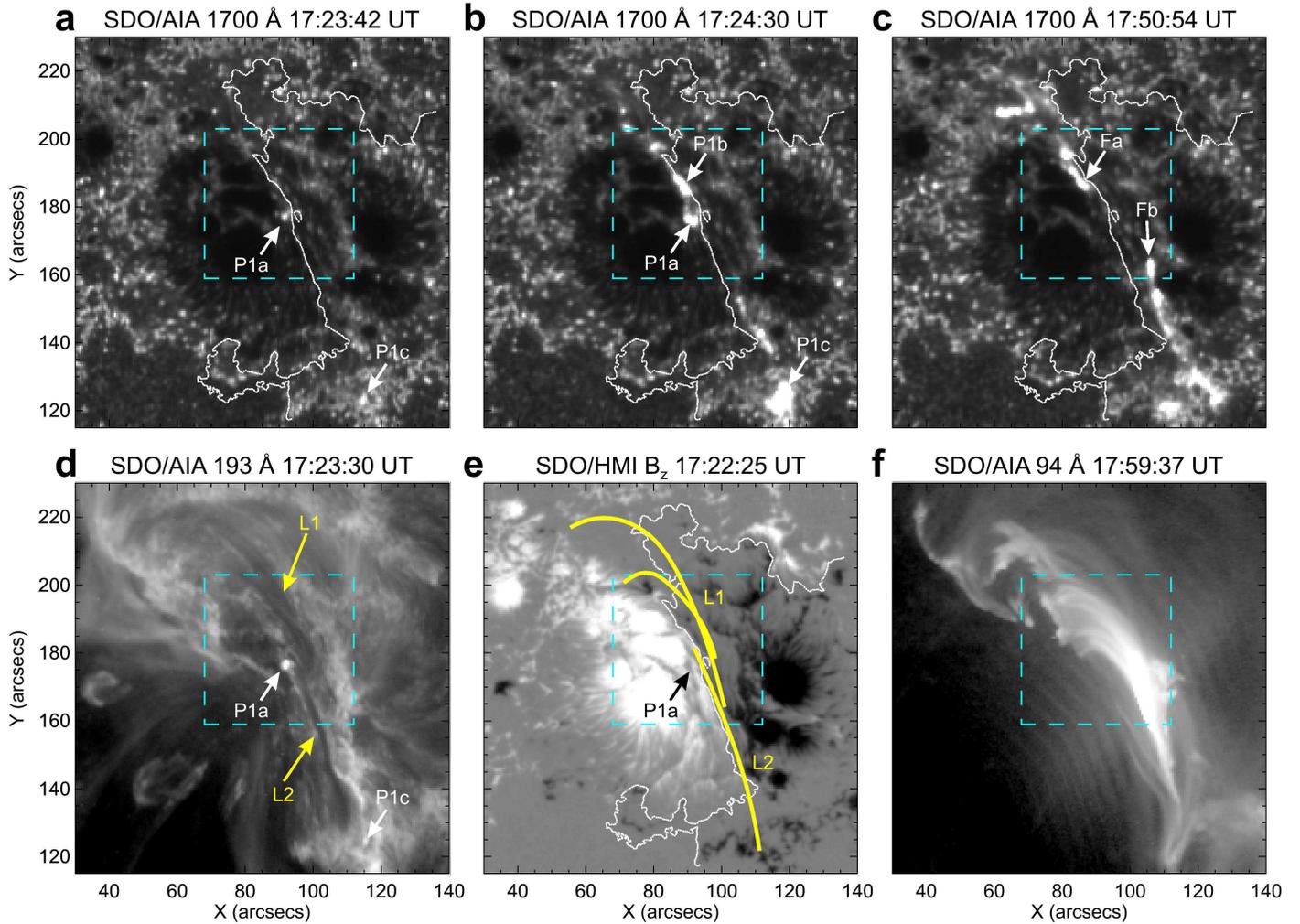}
\vskip3mm
\caption{\textbf{Context images taken by SDO.} (\textbf{a}-\textbf{c}) AIA 1700~\AA\ images at the time of the precursor episode P1 showing flare kernels P1a--P1c, and at the beginning of flare impulsive phase showing two main flare ribbons f1a and f1b (cf. Fig.~1). (\textbf{d}) AIA 193~\AA\ image at P1, clearly showing the kernel P1a lying around the footpoints of large-scale sheared arcades L1 and L2. (\textbf{e}) HMI vertical magnetic field (scaled between $\pm$1500~G), overplotted with yellow lines approximately illustrating the large-scale sheared arcades L1 and L2 seen in \textbf{d} (also consistent with L1 and L2 in Fig.~1f in a smaller FOV). The white curve in \textbf{a}-\textbf{c} and \textbf{e} depicts the main PIL. (\textbf{f}) AIA 94~\AA\ image near the peak of flare nonthermal emission in microwave, showing hot flaring loops. The cyan dashed box denotes the FOV of Fig.~1. \label{sf2}}
\end{figure*}

\begin{figure*}[!t]
\makeatletter
\renewcommand{\fnum@figure}{Supplementary Figure~\thefigure}
\makeatother
\centering
\includegraphics[width=4.6in]{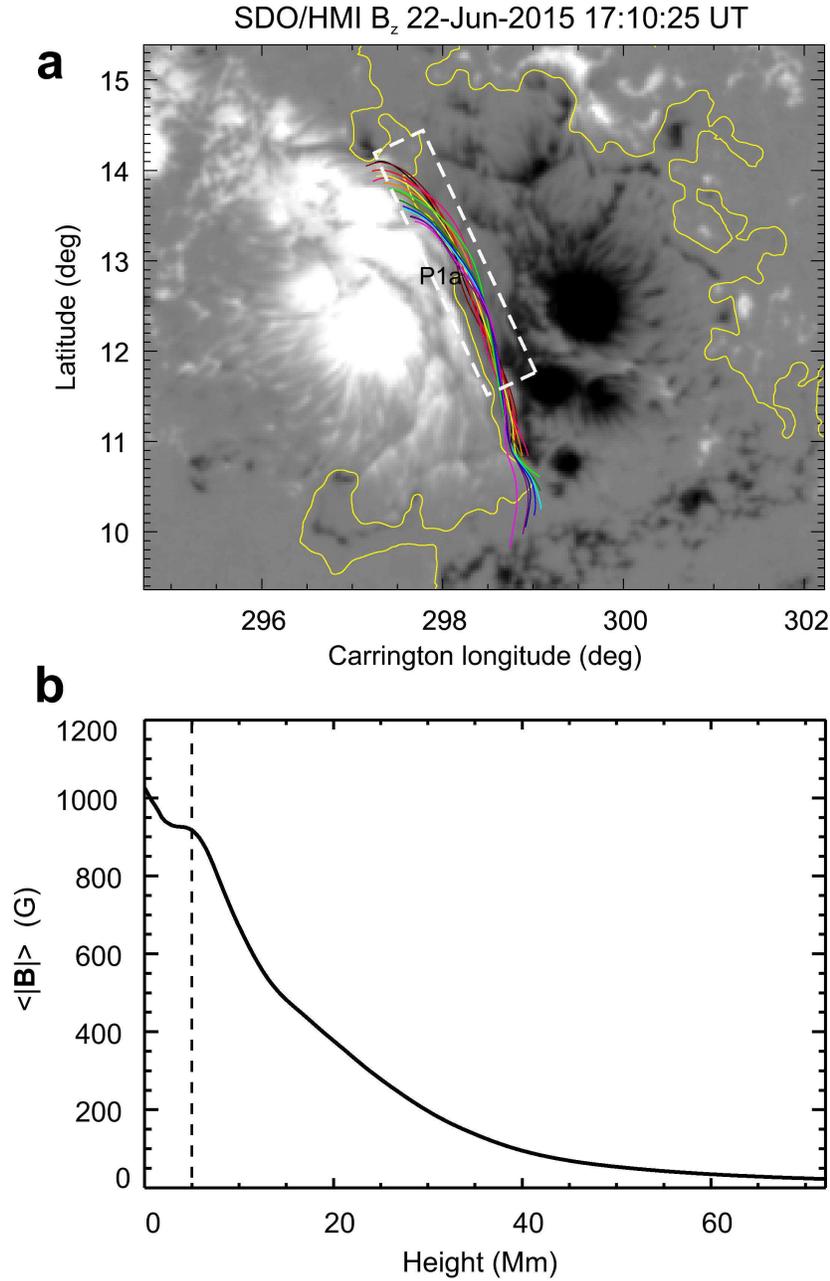}
\vskip2mm
\caption{\textbf{Magnetic field extrapolation model.} (\textbf{a}) SDO/HMI vertical magnetic field at a preflare time overplotted with selected field lines of a twisted flux rope (with an average twist number of $-1.86$ and a s.d. of 0.21) from a NLFFF extrapolation model (see Methods). The location of the precursor brightening kernel P1a (see Fig.~1a) is marked. The yellow curve indicates the main PIL. (\textbf{b}) Average magnetic field strength $\langle |\mathbi{B}| \rangle$ (for the boxed region in \textbf{a} where precursor brightenings are observed with NST) as a function of height $h$, based on the NLFFF model. The error of field strength could be up to about 100~G at the bottom boundary. Note that the bottom boundary ($h=0$) corresponds to the chromospheric level as a result of the preprocessed bottom boundary (see Methods). The vertical dashed line corresponds to 5~Mm, which is the height of the center of the modeled flux rope. \label{sf3}}
\end{figure*}

\begin{figure*}[!t]
\makeatletter
\renewcommand{\fnum@figure}{Supplementary Figure~\thefigure}
\makeatother
\centering
\includegraphics[width=7.5in]{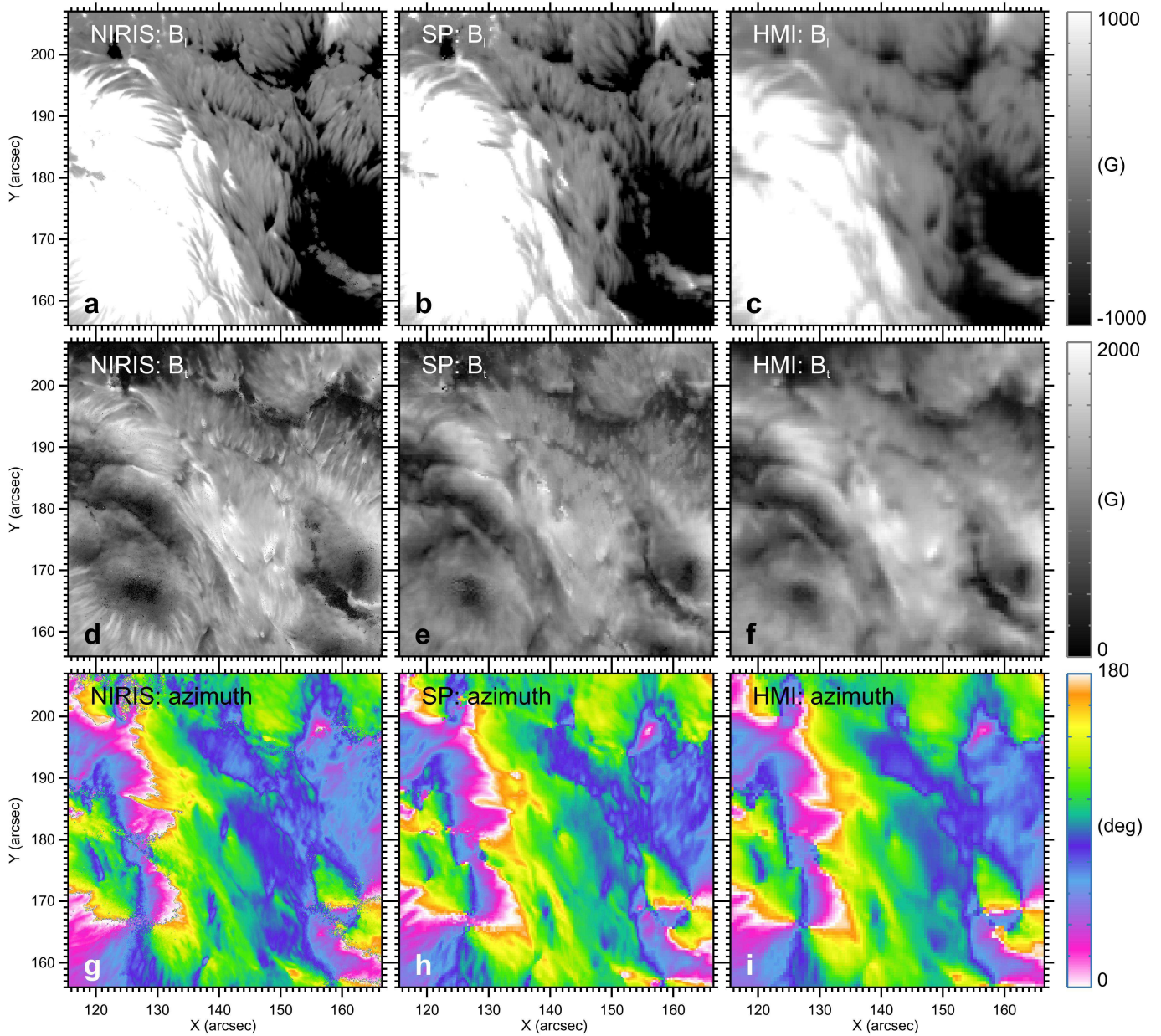}
\vskip1mm
\caption{\textbf{NIRIS magnetogram validation.} Comparison of LOS field $B_l$ (\textbf{a}-\textbf{c}), transverse field $B_t$ (\textbf{d}-\textbf{f}), and the azimuth angle (\textbf{g}-\textbf{i}) among vector magnetograms from NIRIS, Hinode/SP, and SDO/HMI, all taken on 22 June 2015 around 22:35~UT. Data taken by NIRIS matches well with those from Hinode/SP and SDO/HMI, and demonstrates its superiority in high spatiotemporal resolution. See Methods for details.\label{sf4}}
\end{figure*}

\begin{figure*}[!t]
\setcounter{figure}{0}
\makeatletter
\renewcommand{\fnum@figure}{Supplementary Video~\thefigure}
\makeatother
\centering
\includegraphics[width=6in]{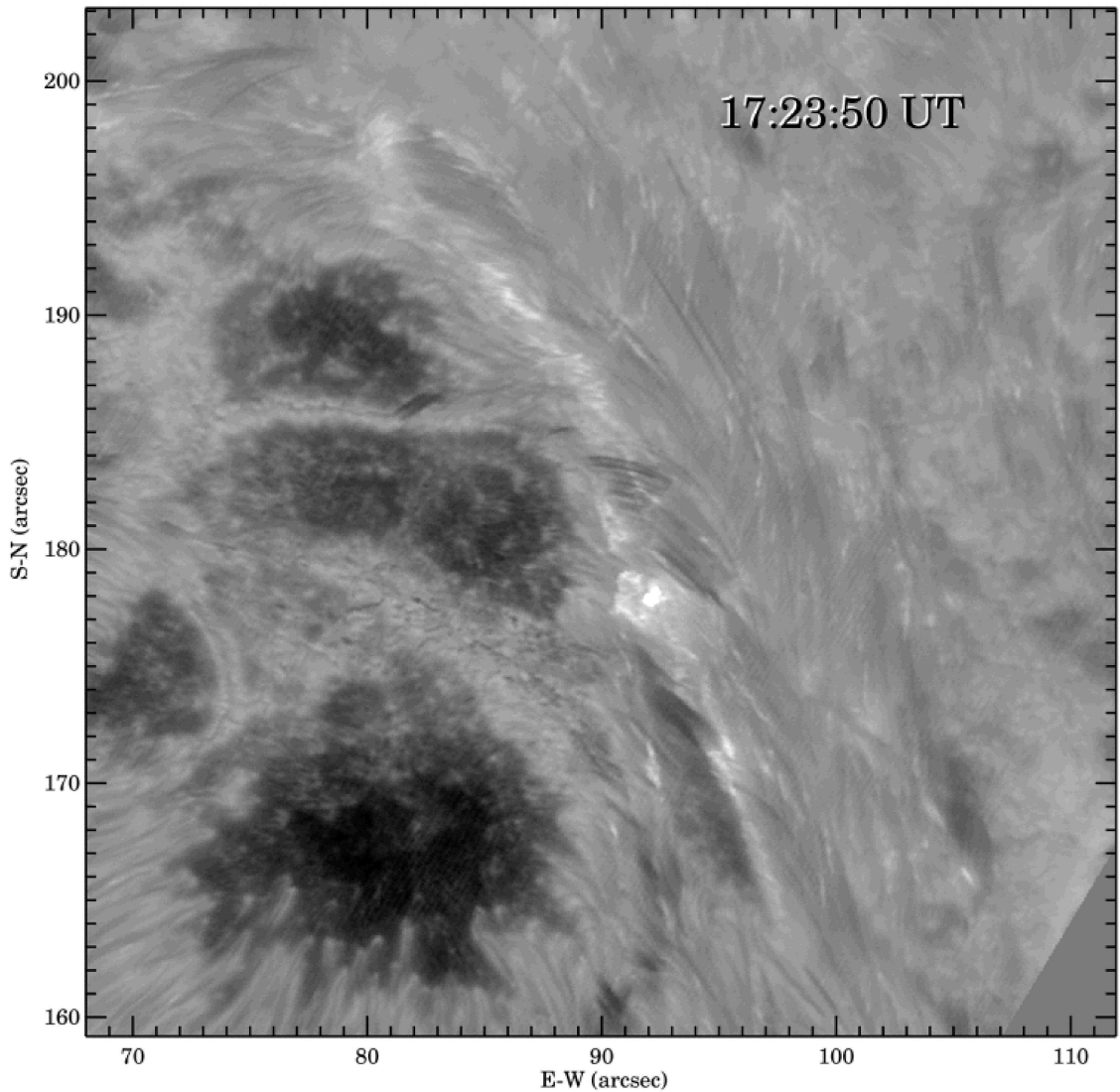}
\caption{\textbf{Time sequence of BBSO/NST \har~\AA\ images}. The video shows the entire evolution of the flare of interest including precursor brightenings and the main impulsive phase (video available from Nature Astronomy web site).}
\end{figure*}

\begin{figure*}[!t]
\makeatletter
\renewcommand{\fnum@figure}{Supplementary Video~\thefigure}
\makeatother
\centering
\includegraphics[width=7.4in]{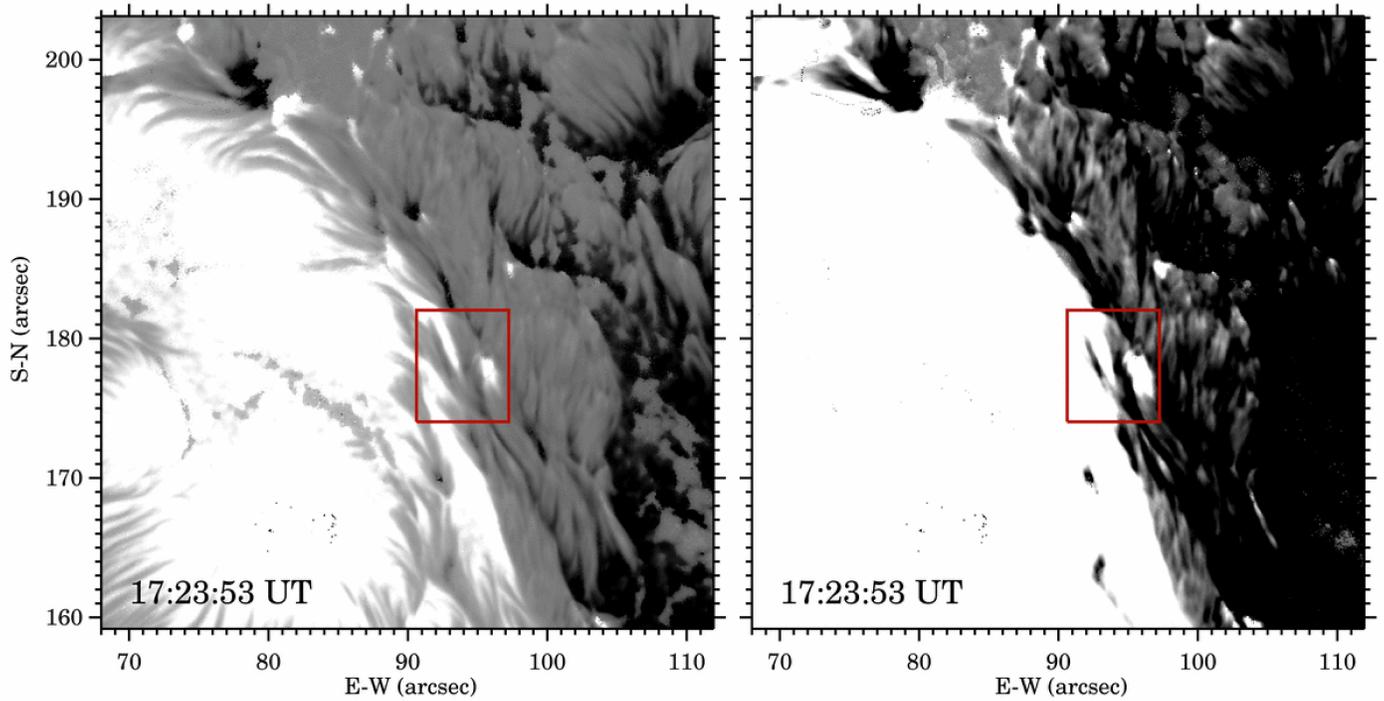}
\vskip2mm
\caption{\textbf{Time sequence of BBSO/NST NIRIS photospheric vertical magnetic field images}. The video has the same FOV as those of Figs~1 and 2a,c. The red box encloses the magnetic channel region, same as that drawn in Fig.~2b for the magnetic flux and electric current calculation. Images in the left panel are scaled between $\pm$1500~G, while those in the right panel are scaled between $\pm$200~G in order to show the magnetic channel structure more clearly (video available from Nature Astronomy web site).}
\end{figure*}

\end{document}